\documentclass[]{aa}
\usepackage[utf8]{inputenc}
\usepackage{graphicx}
\usepackage{txfonts}
\usepackage{natbib}
\bibpunct{(}{)}{;}{a}{}{,}
\graphicspath{{figures/}}
\title{Solar chromosphere heating and generation of plasma outflows by impulsively generated two-fluid Alfv\'en waves}
\authorrunning {M.~Pelekhata, K.~Murawski, S.~Poedts}
\titlerunning {Two-fluid Alfv\'en waves in the solar atmosphere}
   \author{M. Pelekhata\inst{1}
          \and
          K. Murawski\inst{1}
          \and 
          S. Poedts\inst{2,1}
          }

   \institute{Institute of Physics, University of M. Curie-Sk{\l}odowska, 
              Pl.\ M.\ Curie-Sk{\l}odowskiej 1, 20-031 Lublin, Poland\\
         \and
            Centre for Mathematical Plasma Astrophysics / Department of Mathematics, KU Leuven, Celestijnenlaan 200B, 3001 Leuven, Belgium\\
             }

\begin{document}

\abstract
{
We address the heating of the solar chromosphere and the related generation of plasma inflows and outflows.
}
{
We attempt to detect variations in ion temperature and vertical plasma flows, which are driven by impulsively excited two-fluid Alfv\'en waves. We aim to investigate the possible contribution of these waves to solar chromosphere heating and plasma outflows.
}
{
We performed numerical simulations of the generation and evolution of Alfv\'en waves with the use of the JOANNA code, which solves the two-fluid equations for ions+electrons and neutrals, coupled by collision terms.
}
{
We confirm that the damping of impulsively generated small-amplitude Alfv\'en waves slightly affects the temperature of the chromosphere and generates slow plasma flows. In contrast, the Alfv\'en waves generated by large-amplitude pulses increase the chromospheric plasma temperature more significantly and result in faster plasma outflows. The maximum heating occurs when the pulse is launched from the central photosphere, and the magnitude of the related plasma flows grows with  the amplitude of the pulse.
}
{
Large-amplitude two-fluid Alfv\'en waves can contribute significantly to the  heating of the solar chromosphere and to the generation of plasma outflows. 
}

\maketitle

\section{Introduction}
The solar atmosphere can be divided into three layers with different plasma properties: the photosphere (with the conventional surface of a star) extending up to only $500\;$km above the surface, the chromosphere ranging from the top of the photosphere to about $2500\;$km in height, and the solar corona from $2500\;$km and expanding into the solar wind. In the photosphere, the temperature decreases  with height from about $5600\;$K at the bottom until it reaches its minimum of $4300\;$K at the top of the photosphere (or the bottom of the chromosphere), which is\ about $500\;$km above the solar surface \citep{1976ASSL...53.....A}. Higher up, the temperature starts rising again with height, first slowly in the lower chromosphere and then faster in the upper chromosphere until the transition region, which separates the chromosphere from the solar corona. There, the temperature experiences its sudden increase to 1 million K and from there upwards it increases steadily with height in the low corona. A clear explanation of this temperature increase with height above the solar surface remains to be found. As a result of the lower temperatures, the plasma in the low atmospheric layers is only partially ionized \citep{2003ASPC..286..419A}. However, the corona, where the average temperature is about $1 - 3\;$MK, is fully ionized \citep{2005psci.book.....A}. 

Multiple solar missions, such as the Solar Dynamics Observatory (SDO) and the Interface Region Imaging Spectrograph (IRIS), have shown that a diversity of waves occur in the solar atmosphere (\citealt{2009Sci...323.1582J}, \citealt{2011Natur.475..477M}, \citealt{2011ApJ...736L..24O}). The various wave types that occur include Alfv\'en waves \citep{1942Natur.150..405A}. These waves are transverse magnetohydrodynamic (MHD) waves that travel along the magnetic field lines. Alfv\'en waves were reported to be present in both the photosphere and chromosphere (\citealt{2017NatSR...743147S}, \citealt{2021ApJ...907...16B}). As they pass by, they modify the transverse magnetic field and velocity components but, do not alter the gas pressure or the mass density \citep{2005LRSP23N}, at least in the linear limit and in a homogeneous background plasma. A thorough understanding of Alfv\'en waves is essential because they could be a part of the solution to the major problems of heliophysics, such as the solar coronal heating and wind acceleration (\citealt{1974SoPh...35..451U}, \citealt{2010LRSP....7....4O}). Recent theoretical research revealed that Alfv\'en waves can carry enough energy to heat the solar corona \citep{2016ApJ...819L..24Y}. However, the details of the mechanism(s) of the thermal energy release related to their dissipation remain unknown. One potential candidate for that may be associated with ion--neutral collisions \citep{2017ApJ...840...20S}. \citet{1956Obs....76...21P}, \citet{1961ApJ...134..347O}, and \citet{1992Natur.360..241H} were the first to study ion--neutral collisions, but they did not find that this interaction affects the chromospheric temperature. \citet{2018SSRv..214...58B} showed that ambipolar diffusion leads to substantial chromospheric heating, and \citet{2013A&A...549A.113Z} derived a dispersion relation for two-fluid Alfv\'en waves and confirmed that the damping of Alfv\'en waves resulting from the ion--neutral collisions is quite significant. \citet{2017PPCF...59a4038K}, based on a two-fluid model, stated that the presence of neutrals affects the solar atmosphere. The effect of ion--neutral interactions is expected to influence the energy balance of the chromosphere. \citet{2013A&A...549A.113Z} also confirmed that low- and high-frequency photospheric Alfv\'en waves might not reach the solar corona because ion--neutral collisions damp them very efficiently in the upper chromosphere. According to \citet{2011JGRA..116.9104S}, the rate of Alfv\'en wave damping varies with magnetic field strength and wave frequency. For a strong magnetic field, wave damping is low. Low-frequency waves are also weakly damped, and so there is a chance to detect low-frequency Alfv\'en waves in the solar corona under the condition of a strong magnetic field.

According to \citet{1978SoPh...56..305H,1981SoPh...70...25H}, \citet{1982SoPh...75...35H}, \citet{1999ApJ514493K}, and \citet{2012ApJ...749....8M}, Alfv\'en waves can potentially also induce plasma outflows. More recent studies by \citet{2016ApJ...819L..24Y} and  \citet{2017ApJ...840...64S} indeed show that Alfv\'en waves manage to produce fast plasma outflows. As \cite{Tu2005} stated, we can detect outflows in the solar atmosphere between 5 and $20\;$Mm altitude, and at the height of $20\;$Mm, the outflow speed reaches about $10\;\text{km}\cdot \text{s}^{-1}$. A notable problem is to examine such outflows in the lower solar atmosphere, specifically in the chromosphere and transition region \citep{2008ApJ...685.1262M, 2009ApJ...704..883T, 2011Natur.475..477M, 2012SSRv..172...69M, 2015SoPh..290.2889K}. 

The goal of the present paper is to study impulsively generated two-fluid Alfv\'en waves in the partially ionized lower layers of the solar atmosphere. These studies are performed numerically in the context of the observed atmospheric heating and plasma outflows, which in the higher coronal regions may result in nascent solar wind.

The organisation of the remainder of this paper is as follows. Section~2 presents the governing two-fluid equations,  an initial equilibrium model of the solar atmosphere, and the impulsive perturbations we applied in the simulations. In Section~3, the results of the numerical simulations are presented and Section~4 contains the conclusions of this paper with a discussion and summary of the results of the numerical experiments performed.
\section{Two-fluid numerical model of the partially ionized solar atmosphere}
In the dens(er) lower layers of the solar atmosphere, the interactions between the different plasma components are commonly considered to be collective, which means that each particle simultaneously affects several other particles. Therefore, it is justified to consider the plasma in these atmospheric layers as a continuum, that is, as a fluid \citep{2002PhPl....9.4052R}. The temperature depends on the altitude of the considered layer in the solar atmosphere. There, where the temperature is high enough, as in the transition layer and the corona mentioned above, the plasma is fully ionized. However, as mentioned in the introduction, in the photosphere and chromosphere the temperatures are lower and the plasma there is not fully ionized. In the photosphere, the ionization degree is even as low as 0.01\%. As discussed above, it is therefore necessary to take into account the dynamics of neutral particles in these atmospheric layers. As a simple conceivable approach, we use the two-fluid plasma model to describe the partially ionized atmosphere, which consists of ions+electrons and neutrals, treated as two separate fluids \citep{2011A&A...529A..82Z}. Each of these two fluids has its own mass density, flow velocity, and gas pressure and the fluids interact with each other through ion--neutral collisions.
\subsection{Two-fluid equations}
The two-fluid approach combines the fluid mechanics theory (governed by the Navier-Stokes equations) and electromagnetism (governed by Maxwell' s equations). The evolution equations in the framework of a two-fluid model can therefore be written in the following form \citep{2018SSRv..214...58B, 2019ApJ...884..127W}:
\begin{eqnarray}
    &&\frac{\partial \varrho_{\rm i}} {\partial t}+\nabla\cdot(\varrho_{\rm i} \mathbf{V}_{\rm i}) = 0\,, 
    \label{eq:ion_continuity} \\
    &&\frac{\partial \varrho_{\rm n}}{\partial t}+ \nabla\cdot(\varrho_{\rm n} \mathbf{V}_{\rm n}) = 0\,, 
    \label{eq:neutral_continuity} \\
    &&\frac{\partial (\varrho_{\rm i} \mathbf{V}_{\rm i})}{\partial t} + \nabla \cdot (\varrho_{\rm i} \mathbf{V}_{\rm i} \mathbf{V}_{\rm i} + p_{\rm i e}\mathbf{I}) = \\ \nonumber && \varrho_{\rm i} \mathbf{g} + \frac{1}{\mu}(\nabla \times \mathbf{B}) \times \mathbf{B} - \mathbf{S_{\rm m}}\,,
    \label{eq:ion_momentum}\\
    &&\frac{\partial (\varrho_{\rm n} \mathbf{V}_{\rm n})}{\partial t} + \nabla \cdot (\varrho_{\rm n} \mathbf{V}_{\rm n} \mathbf{V}_{\rm n} + p_{\rm n} \mathbf{I})  =  \varrho_{\rm n} \mathbf{g} + \mathbf{S_{\rm m}}\,,
    \label{eq:neutral_momentum} \\
    &&\frac{\partial E_{\rm i}}{\partial t} + \nabla\cdot\left[\left(E_{\rm i}+p_{\rm i e} + \frac{\mathbf{B}^2}{2\mu} \right)\mathbf{V}_{\rm i}-\frac{\mathbf{B}}{\mu}(\mathbf{V}_{\rm i}\cdot \mathbf{B})\right] = \\ \nonumber && (\varrho_{\rm i} \mathbf{g} + \mathbf{S_{\rm m}}) \cdot \mathbf{V}_{\rm i} + Q_{\rm i}\,, 
    \label{eq:ion_energy} \\
   &&\frac{\partial E_{\rm n}}{\partial t}+\nabla\cdot[(E_{\rm n}+p_{\rm n})\mathbf{V}_{\rm n}] = (\varrho_{\rm n} \mathbf{g} + \mathbf{S_{\rm m}})\cdot\mathbf{V}_{\rm n} + Q_{\rm n}\,,
    \label{eq:neutral_energy} \\
\hbox{with}&& \nonumber\\
    &&E_{\rm i} = \frac{\varrho_{\rm i}\mathbf{V}_{\rm i}^2}{2} + \frac{p_{\rm i e}}{\gamma -1 } + \frac{{\mathbf B}^2}{2\mu}\,, 
    \hspace{3mm} 
%
    E_{\rm n} = \frac{\varrho_{\rm n}\mathbf{V}_{\rm n}^2}{2} + \frac{p_{\rm n}}{\gamma -1 }\,,  \\
    &&\frac{\partial \mathbf{B}}{\partial t} = \nabla \times (\mathbf{V_{\rm i} \times }\mathbf{B}),
    \hspace{6mm}
    \nabla \cdot{\mathbf B} = 0\,,
    \label{eq:ions_induction} \\
\hbox{where}&& \nonumber\\
    &&p_{\rm ie}=\frac{k_{\rm B}}{m_{\rm H}\mu_{\rm i}}\varrho_{\rm i} T_{\rm i}\,,
    \quad
   p_{\rm n}=\frac{k_{\rm B}}{m_{\rm H}\mu_{\rm n}}\varrho_{\rm n} T_{\rm n}\,. \\
   &&\mathbf{S_{\rm m}}=v_{\rm in}\varrho_{\rm i}(\vec{V_{\rm i-}}\vec{V_{\rm n}})\,, \\
    &&Q_{\rm i}=\frac{1}{2}\nu_{\rm in}\varrho_{\rm i}(\vec{V_{\rm i}}-\vec{V_{\rm n}})^2-\frac{\nu_{\rm in}\varrho_{\rm i}k_{\rm B}}{(\gamma-1)m_{\rm H}\mu_{\rm n}}(T_{\rm i}-T_{\rm n})\,,  \\
    &&Q_{\rm n}=\frac{1}{2}\nu_{\rm in}\varrho_{\rm i}(\vec{V_{\rm i}}-\vec{V_{\rm n}})^2-\frac{\nu_{\rm in}\varrho_{\rm i}k_{\rm B}}{(\gamma-1)m_{\rm H}\mu_{\rm n}}(T_{\rm n}-T_{\rm i})\,.
\end{eqnarray}
Here, $\varrho_{\rm i,n}$ is the mass density of the ions and neutrals, respectively, and similarly $\vec{V}_{\rm i,n}$ are the velocity fields, $p_{\rm i,n}$ the gas pressures, $E_{\rm i,n}$ the total energy densities, while $\vec{g} = [0,-g,0]$ denotes the gravitational acceleration with its magnitude $g=274.78\;\text{m}\cdot\text{s}^{-2}$. Moreover, $k_{\rm B}$ is the Boltzmann constant, $\vec{I}$ denotes the identity matrix, $\mu$ the magnetic permeability, $\vec{B}$ the magnetic field, $\gamma = 1.4$ the adiabatic coefficient, and $T_{\rm i,n}$ corresponds to the temperatures of the two fluids. Also, $m_{\rm H}$ is the mass of a hydrogen atom, and $\mu_{\rm i} = 0.58, \, \mu_{\rm n} = 1.21$ denote the mean mass of each species. Hence, the indices $\rm_{i, e, n}$ correspond respectively to ions, electrons, and neutrals, while $\nu_{\rm i,n}$ is the  ion--neutral collision frequency which is given as \citep{1965RvPP....1..205B, 2018SSRv..214...58B}
    \begin{equation}
        \nu_{\rm in}=\frac{4}{3}\frac{\sigma_{\rm in}\varrho_{\rm n}}{m_{\rm H}(\mu_{\rm i}+\mu_{\rm n})}\sqrt{\frac{8k_{\rm B}}{\pi m_{\rm H}}\left(\frac{T_{\rm i}}{\mu_{\rm i}}+\frac{T_{\rm n}}{\mu_{\rm n}}\right)}\, ,
    \end{equation}
with $\sigma_{\rm in}$ being the collision cross-section, which varies with the energy of the colliding particles, that is, with temperature. Here, we adopt its 
classical value, $\sigma_{\rm in}=1.4\times 10^{-19}$~m$^{2}$, 
\citep{VranjesKrstic2013}. 
\subsection{Magnetohydrostatic equilibrium}
The solar corona continuously expands into interplanetary space. 
Yet, for computational economy during numerical simulations, we assume that the solar atmosphere is in hydrostatic equilibrium. This is the state in which the radially inward force of gravity and the radially outward gas pressure gradient force are in balance. This state is described by the hydrostatic condition, expressing this force balance, that is,
    \begin{equation}
        -\nabla p_{\rm i,n}+\varrho_{\rm i,n}\vec{g}=\bf0\,.
    \end{equation}

We initialise the velocity, temperature, pressure, and mass density of both plasma components from the hydrostatic equilibrium state given by the solution of Eq.~(14). This means that the velocity of both ions $\vec{V_{\rm i}}$ and neutrals $\vec{V_{\rm n}}$ is set equal to zero. Additionally, we set the initial temperature to be equal for both plasma components, $T_{\rm i}=T_{\rm n}=T$, according to the semi-empirical model of \citet{2008ApJS..175..229A}. The following equations describe the initial equilibrium values of gas pressure and mass density \citep[e.g.,][]{2017ApJ...849...78K}:
\begin{eqnarray}
       && p_{\rm i,n}(y)=p_{\rm 0i,n} \exp \left(-\int_{y_{\rm r}}^{y}	\frac{dy'}{\Lambda(y' )}\right)\,,\\
\hbox{and}&& \nonumber\\%
        &&\varrho_{\rm i,n}(y)=\frac{p_{\rm i,n}(y)}{g\Lambda_{\rm i,n}}\,,
        \hspace{10mm}
        \varrho_{\rm e}(y)=0\,,\\
\hbox{with}&& \nonumber\\
       && \Lambda_{\rm n}=\frac{k_{\rm B} T_{\rm n}}{g\mu_{\rm n} m_{\rm H}}\,,
        \hspace{10mm}
        \Lambda_{\rm i}=\frac{k_{\rm B} T_{\rm i}}{g\mu_{\rm i} m_{\rm H}}\,.
    \end{eqnarray}
Here, $\Lambda_{\rm in}$ denote the ion and neutral pressure scale heights, while $p_{\rm 0in}$ are the plasma and gas pressures at the reference height, $y_{\rm r}$ is the reference height taken as $50\;$Mm, with $p_{\rm 0i}=10^{-2}$~Pa and $p_{\rm 0n}=3\cdot 10^{-4}$~Pa, and $m_{\rm H}$ corresponds to the mass of a hydrogen atom. 

We also assume that a magnetic field penetrates this hydrostatic equilibrium state, but we consider a magnetic field that is force-free $(\text{i.e.,}\ (\nabla\times\vec{B})\times\vec{B}/\mu=\vec{0})$ and even current-free $(\text{i.e.,}\ (\nabla\times\vec{B})/\mu=\vec{0})$, so that the hydrostatic force balance is not disturbed. A uniform vertical magnetic field $\vec{B}=B_{\rm y}{\vec{\hat y}}$, with a magnitude $B_{\rm y}$ that is chosen to be equal to $30\;$G, satisfies these conditions or equations. Here, ${\vec{\hat y}}$ denotes a unit vector directed along the $y$-axis. The magnetohydrostatic equilibrium model assumes a uniform and vertical magnetic field, while in the real solar atmosphere the magnetic field expands with height. Recent works that include such magnetic field line expansion include those by \citet{2017ApJ...840...20S, 2019ApJ...871....3S}. This expansion is more important in the low chromosphere and our model requires a revision there.

Figure~1 (top panel) illustrates the vertical variation of the initial equilibrium temperature $T(y)$. This plot reveals that the minimum temperature is located $100\;$km above the photosphere (occupying the region given by $0.0\;$Mm $\le y \le 0.5\;$Mm), mainly at $y = 0.6\;$Mm, and $T(y)$ becomes slightly higher in the middle and upper chromosphere $(0.5\;$Mm $\le y \le 2.1\;$Mm). A sudden increase in temperature occurs at the height corresponding to the transition region $(y \approx 2.1\;$Mm). Above the transition region, in the solar corona, the temperature keeps rising with height until it reaches a magnitude of $1\;$MK at $y = 20\;$Mm.

The bottom panel of Figure~1 displays the vertical variation of the Alfv\'en speed $c_{\rm a}$, which is defined as
    \begin{equation}
        c_{\rm a}=\frac{B_{\rm y}}{\sqrt{\mu(\varrho_{\rm 0i} + \varrho_{\rm 0n})}}\,.
    \end{equation}
    
At the bottom of the photosphere, located at $y = 0\;$Mm, $c_{\rm a} \approx 200\;\text{m}\cdot\text{s}^{-1}$. We note that $c_{\rm a}$ grows with height and the most drastic growth takes place at the transition region. However, in the solar corona, $c_{\rm a}(y=20\; \text{Mm}) \approx 10^6\;\text{m}\cdot\text{s}^{-1}$, and it grows steadily with increasing $y$ (height).

\begin{figure}
    \begin{center}
        \hspace{-0.2cm}
        \includegraphics[width=0.4\textwidth]{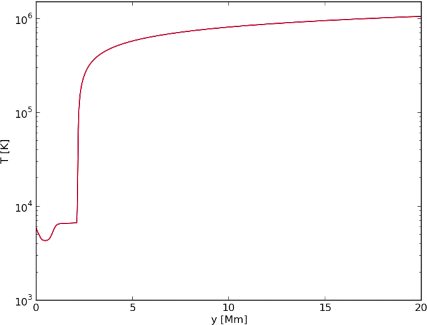}
        \vspace{0.5 cm}
        \includegraphics[width=0.4\textwidth]{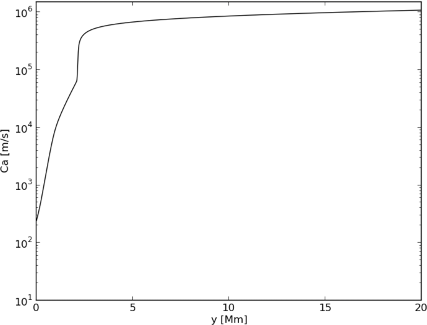}
    \end{center}
    \vspace{-0.5cm}
    \caption{Variation with height $y$ of the equilibrium temperature $T$, given by \protect\cite{2008ApJS..175..229A}, (top) and Alfv\'en speed (bottom).}
    \label{fig:profiles}
\end{figure}
\subsection{Impulsive perturbations}
In order to generate Alfv\'en waves, we perturb the magnetohydrostatic equilibrium by initially (i.e.\ at $t = 0\;$s) launching a localized pulse in the transverse component of ion velocity, $V_{\rm iz}$. The following equation describes this pulse: 
    \begin{equation}
      V_{\rm iz}(y, t=0\;\text{s})=A\; \exp\left(-\frac{(y-y_{\rm 0})^2}{w^2}\right)\,.
    \end{equation}
Here, $A$ denotes the amplitude of the pulse, $w$ its width, and $y_{\rm 0}$ its vertical location. By varying $A$ and $y_{\rm 0}$, but holding the width $w$ fixed at $0.2\;$Mm, a similar pulse of different amplitude $A$ is launched in the photosphere or chromosphere.
\section{Numerical results}
In order to study Alfv\'en waves in the inhomogeneous and only partially ionized lower layers of the solar atmosphere, we have to rely on numerical simulations. In our model, initially (at $t = 0\;$s), an ion velocity pulse described by Eq.~(19) is injected into the simulation domain specified as $(-0.08, 0.08) \times (-0.5, 60)\;$Mm$^2$. We use the JOANNA code \citep{Wojciketal2019a} to perform the numerical simulations, that is, to solve the two-fluid equations presented and discussed in the previous section. Indeed, this code solves the initial-boundary value problem for the specified two-fluid model. All plasma variables are initialised with their magnetohydrostatic values at the lower and upper boundaries of the simulation domain and held fixed in time throughout the simulation. Setting  non-reflecting or absorbing boundary conditions along the gravity action is a formidable and still not fully solved task. We have implied the simplest conceivable boundary conditions with all plasma quantities set to their equilibrium values. Such conditions being supplemented by a well stretched grid along the $y$-direction is found to significantly reduce reflections of the incoming signal. The part of the simulation domain specified by $-0.5\; \text{Mm} \le y \le 4.62\;$Mm is divided into 512 grid cells, leading to a numerical grid cell size $\Delta y = 10\;$km in that area. The region higher up is covered by 128 grid cells and in this part of the computational domain the grid is stretched, that is, the size of the grid cells steadily grows with height. At the side ($x-$) boundaries, so-called "open" boundary conditions are implemented, meaning that the $x$-derivatives of all the plasma quantities are set to zero.
\subsection{Numerical verification test}
Before starting any simulation, we first verify the numerical accuracy by trying to quantify the numerical errors. One test consists of running the code without any initial pulse and relaxing the initial (analytical) equilibrium state numerically, that is, allowing the physical system to evolve in time without any external influence. Although a zero–amplitude pulse cannot lead to the generation of Alfv\'en waves, as a result of the numerical approximation which is due to the discretization of the equations on a mesh with finite resolution, there is some signal present in the vertical ($y-$) component of the ion velocity and temperature in the system. This signal corresponds to magnetoacoustic waves, which are affected by gravity. We can trace these waves by plotting the $V_{\rm iy}$ component in the $(y,t)$plane (Fig.~2, top) and also plotting the relative perturbed ion temperature, 
    \begin{equation}
      \frac {\delta T_{\rm i}}{T} = \frac{T_{\rm i} - T}{T}\,;
    \end{equation}
see the bottom plot of Fig.~2. We note that the obtained max ($V_{\rm iy}) \approx 0.34\;\text{km}\cdot\text{s}^{-1}$, which takes place in the corona at $y \approx 20\;$Mm, and it is very small compared to the local sound speed $c_{\rm s} \approx 100\;\text{km}\cdot\text{s}^{-1}$ there (not shown). The signal propagates upward at an average velocity of $125\;\text{km}\cdot\text{s}^{-1}$, which corresponds to the average sound speed (the average Alfv\'en velocity in that area is an order of magnitude higher). Similarly, max ($\delta T_{\rm i}/T) = 0.0045$, which is also a very small perturbation and it propagates upward at the same velocity as the $V_{\rm iy}$ perturbation. Hence, we infer that for the chosen numerical grid and discretization scheme, the generated numerical noise in the relaxation phase is rather small and does not affect the numerical results presented below. 
\begin{figure}
    \begin{center}
        \hspace{-0cm}
        \includegraphics[width=0.4\textwidth]{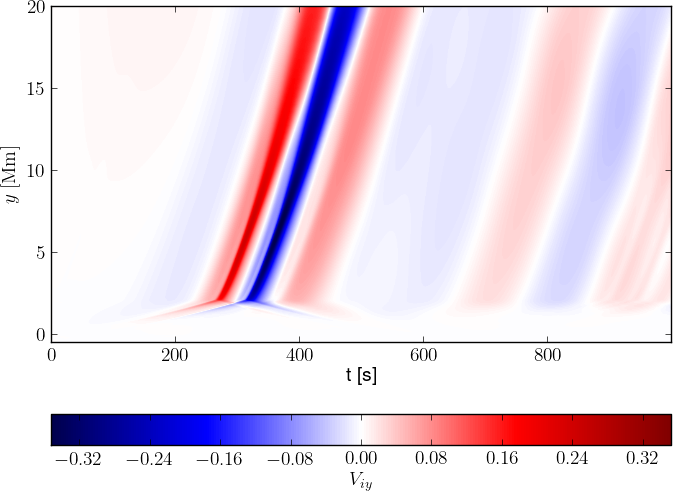}
        \hspace{0cm}
        \includegraphics[width=0.4\textwidth]{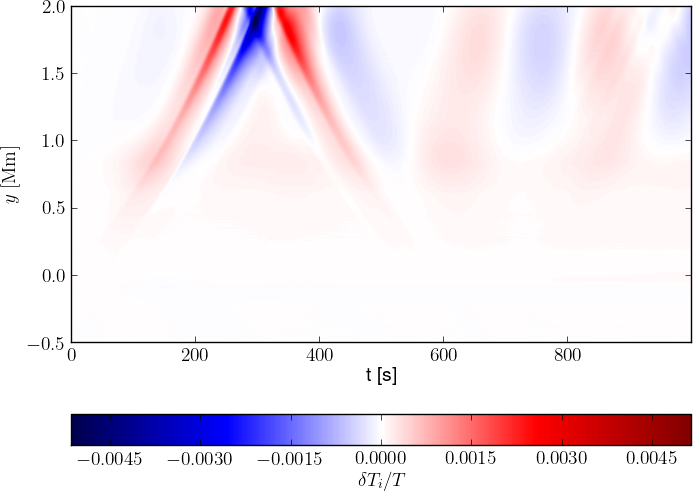}
    \end{center}
    \vspace{-0cm}
    \caption{Time–distance plots for $V_{\rm iy}$ (top) and $\delta T_{\rm i}/T $ (bottom) for the initial pulse-free system, i.e.,\ $A=0\;\text{km}\cdot\text{s}^{-1}$. In other words, this plot shows the numerical relaxation of the initial equilibrium state.}
    \label{fig:no_pulse}
\end{figure}
\begin{figure*}
    \begin{center}
        \hspace{-0cm}
        \includegraphics[width=0.4\textwidth]{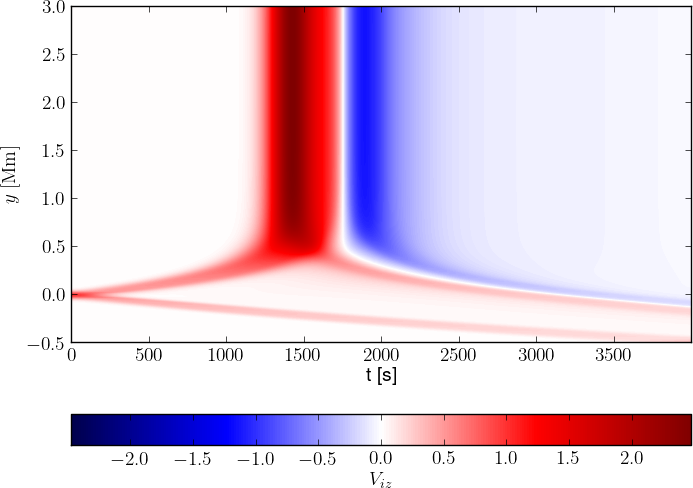}
        \hspace{0cm}
        \includegraphics[width=0.4\textwidth]{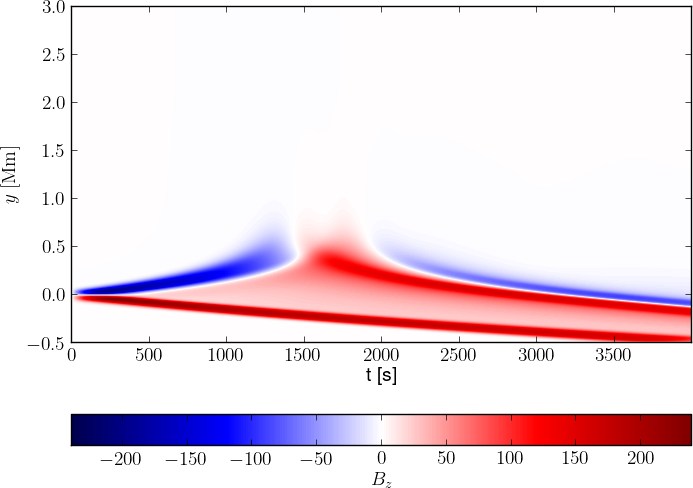}
        \vspace{0cm}
        \includegraphics[width=0.4\textwidth]{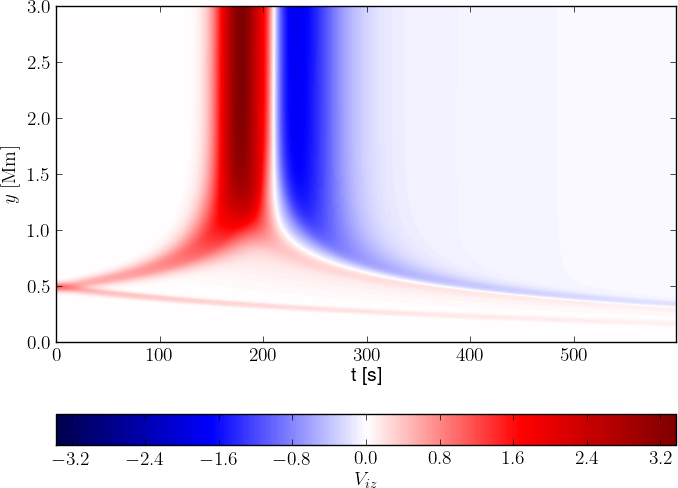}
        \hspace{0cm}
        \includegraphics[width=0.4\textwidth]{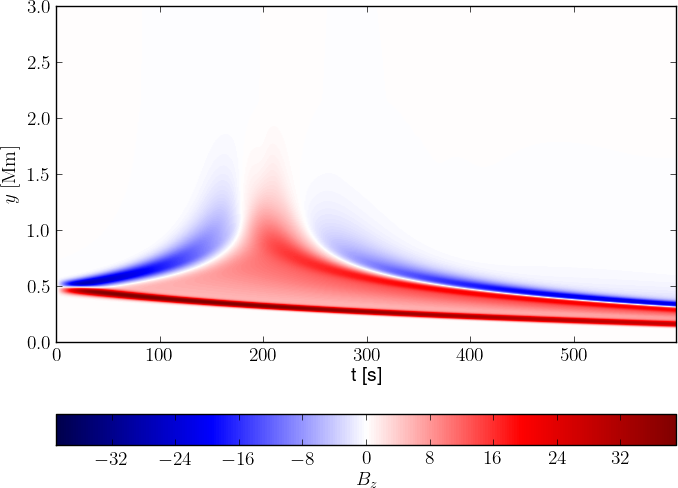}
    \end{center}
    \vspace{-0cm}
    \caption{Time–distance plots for $V_{\rm iz}$ (left) and $B_{\rm z}$ (right) for $A=1\;\text{km}\cdot\text{s}^{-1}$, $y_{\rm 0} = 0\;$Mm (top) and $y_{\rm 0} = 0.5\;$Mm (bottom).}
    \label{fig:maxViz_pulse_1}
\end{figure*}
\subsection{Small-amplitude Alfv\'en waves}
In this section, we consider small-amplitude Alfv\'en waves generated by an initial pulse. Figure~3 displays the evolution of the Alfv\'en waves that are excited by the initial pulse with an amplitude as small as $A=1\;\text{km}\cdot\text{s}^{-1}$ and launched from the bottom of the photosphere, that is, $y_{\rm 0}=0\;$Mm (top plots), and at the top of the photosphere, at $y_{\rm 0}=0.5\;$Mm (bottom plots). The initial pulse splits into two counter-propagating waves, which are damped. This damping results from collisions between ions and neutrals \citep{2005A&A...442.1091L, 2016ApJ...817...94A, {2016ApJ...819L..11S}, {refId0}, {refId02}, {2007A&A...461..731F}, {2019ApJ...871....3S}}. For the initial pulse at the bottom of the photosphere, the upwardly propagating waves experience partial reflection at $y \approx 0.5\;$Mm (upper plots), corresponding to the top of the photosphere. However, when the initial pulse is located at the top of the photosphere, the reflection takes place at about $y \approx 1.0\;$Mm (bottom plots).

In the case of $y_{\rm 0}=0\;$Mm (Fig.~3, left-top), the reflection takes place after $t\approx1500\;$s, so the upward propagating waves travel at about $0.33\;\text{km}\cdot\text{s}^{-1}$, which is the average Alfv\'en velocity in the photosphere (see Fig.~1, bottom). However, for $y_{\rm 0}=0.5\;$Mm (Fig.~3, left-bottom), as a result of higher values of $c_{\rm a} (y)$ in the low chromosphere (Fig.~1, bottom), the waves propagate upward much faster and the reflection therefore occurs much earlier, that is, at about $t\approx200\;$s, which means that the waves indeed propagate with an average velocity of $2.5\;\text{km}\cdot\text{s}^{-1}$. Some waves propagate higher up, reaching the transition region and the solar corona (left). In the photosphere and the low chromosphere, the amplitudes of the upward- and downward-propagating waves decay with the distance of their propagation. This  is a result of ion--neutral collisions taking place in these layers. The left panels of Fig.~3 show that the maximum value of the transversal velocity component is smaller for $y_{\rm 0} = 0\;$Mm with max$(V_{\rm iz}) \approx 2.5\;\text{km}\cdot\text{s}^{-1}$ than for $y_{\rm 0} = 0.5\;$Mm with max$(V_{\rm iz}) \approx 3.4\;\text{km}\cdot\text{s}^{-1}$. These maxima are reached in the corona as the plasma density is much lower there. 

The right panels of Fig.~3 present plots of a transversal magnetic field $B_{\rm z}$, which is in phase with $V_{\rm iz}$ for the downward-propagating Alfv\'en waves and in phase opposition for the upward propagating waves. Indeed, here we also observe upward- and downward-propagating waves from the pulse height. The upward-propagating waves experience reflection in both cases (i.e.\ for $y_{\rm 0}=0\;$Mm and $y_{\rm 0}=0.5\;$Mm) after passing a distance of about $0.5\;$Mm, essentially simultaneously with the reflections seen in the $V_{\rm iz}$ component (Fig.~3, left panels). However, we note that the $B_{\rm z}$ signals evolve differently from the $V_{\rm z}$ perturbations. The $B_{\rm z}$ perturbation amplitude strongly falls off with height and even becomes evanescent above the reflection height, while the $V_{\rm iz}$ component reaches its maximum values there. This difference is well known in the MHD framework, where the evolution of the $B_{\rm z}$ component is described by the following wave equation:
\begin{equation}
    \frac{\partial^2 B_{\rm z}}{\partial t^2}-\frac{\partial}{\partial y}\left(c_a^2\frac{\partial B_{\rm z}}{\partial y}\right)=0\,,
\end{equation}
while the transversal component of velocity, $V_{\rm z}$, is governed by the classic wave equation \citep[e.g.,][]{2017SoPh..292...31W}: 
\begin{equation}
    \frac{\partial^2 V_{\rm z}}{\partial t^2}-c_a^2\frac{\partial^2 V_{\rm z}}{\partial y^2}=0\,.
\end{equation}

Hence, the increase in the Alfv\'en velocity $c_{\rm a}$ with height $y$ yields a relatively large damping term ($\sim {\partial B_{\rm z}}/{\partial y}$) in the wave equation for $B_{\rm z}$ which does not appear in the wave equation for $V_{\rm z}$.

\begin{figure*}
    \begin{center}
        \hspace{-0cm}
        \includegraphics[width=0.4\textwidth]{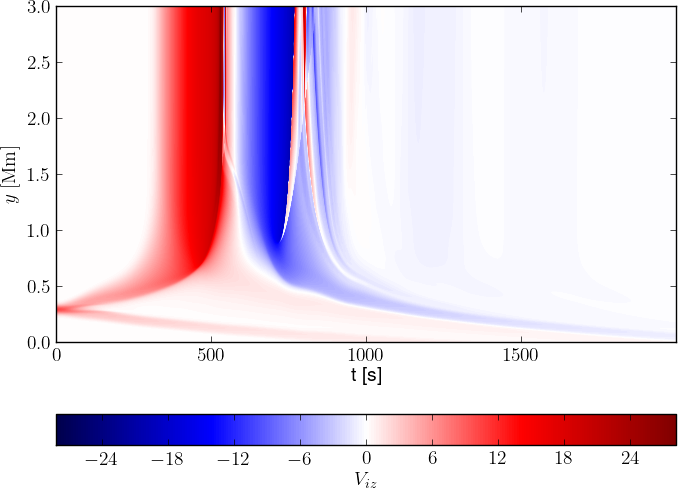}
        \hspace{0cm}
        \includegraphics[width=0.4\textwidth]{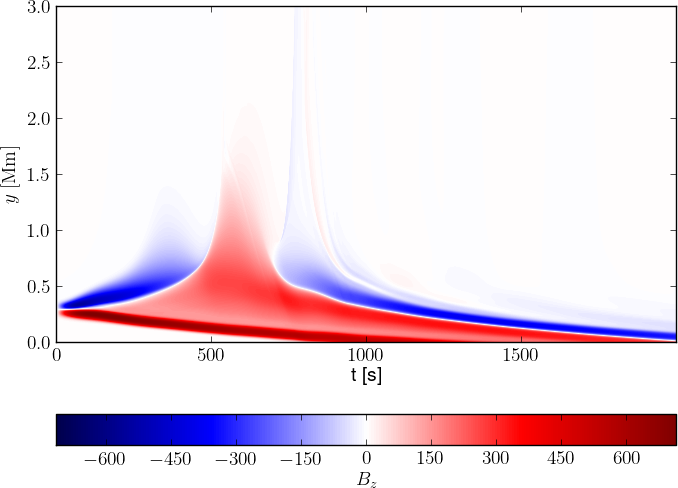}
        \vspace{0cm}
        \includegraphics[width=0.4\textwidth]{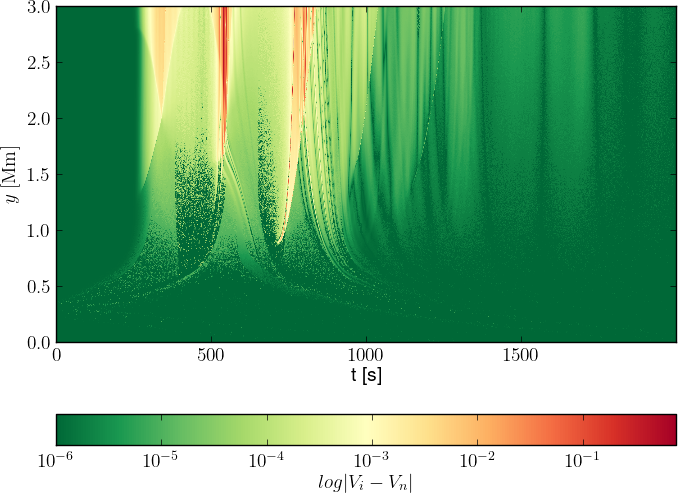}
        \vspace{0cm}
        \includegraphics[width=0.4\textwidth]{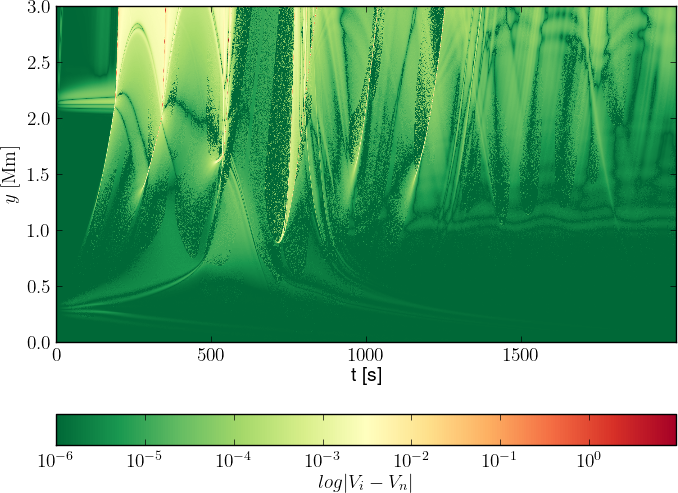}
    \end{center}
    \vspace{-0cm}
    \caption{Time–distance plots for: $V_{\rm iz}$ (top-left), $B_{\rm z}$ (top-right), $V_{\rm iz}-V_{\rm nz}$ (bottom-left), and $V_{\rm iy}-V_{\rm ny}$ (bottom-right) in the case of $A=10\;\text{km}\cdot\text{s}^{-1}$ and $y_{\rm 0} = 0.3\;$Mm.}
    \label{fig:maxViz_pulse_1}
\end{figure*}
\subsection{Heating and plasma outflows by large-amplitude Alfv\'en waves}
In this section, we explore the dynamics of large-amplitude Alfv\'en waves generated by an initial pulse in the lower atmospheric layers of the solar atmosphere. We limit our discussion to the case of an initial pulse of $A=10\;\text{km}\cdot\text{s}^{-1}$. The time--distance profiles of $V_{\rm iz}$ and $B_{\rm z}$ are displayed in Fig.~4 (top), and they globally look similar to those illustrated in Fig.~3 which corresponds to $A=1\;\text{km}\cdot\text{s}^{-1}$. We note that the magnetic field perturbations now travel higher up in the solar atmosphere before being dissipated. The bottom panels of Fig.~4 show the corresponding velocity drifts: $V_{\rm iz}-V_{\rm nz}$ (bottom-left) and $V_{\rm iy}-V_{\rm ny}$ (bottom right). In both bottom panels, it can be seen that the velocity drifts grow with height towards maximum values of $V_{\rm iz}-V_{\rm nz} \approx 1\;\text{km}\cdot\text{s}^{-1}$ and $\max(V_{\rm iy}-V_{\rm ny}) \approx 10\;\text{km}\cdot\text{s}^{-1}$, respectively. Due to the fact that the $\max(V_{\rm iy}-V_{\rm ny})$ value is higher than the $\max(V_{\rm iz}-V_{\rm nz})$ value, we infer that magnetoacoustic waves, which are driven by the Alfv\'en waves, are responsible for the observed heating, which is described by the first term of the right-hand side of Eqs.~(11) and (12). The nonlinear coupling of Alfv\'en waves with magnetoacoustic waves is due to the ponderomotive force. A study of the ponderomotive force in the multi-fluid framework was recently presented by \cite{2018ApJ...856...16M}, who investigated impulsively generated Alfvén waves. On the other hand, the fact that the nonlinearly driven magnetoacoustic waves heat the plasma more efficiently has already been discussed; namely by \cite{2016ApJ...817...94A}, who used the single-fluid approximation.
 
 \begin{figure*}
    \begin{center}
        \hspace{-0cm}
        \includegraphics[width=0.4\textwidth]{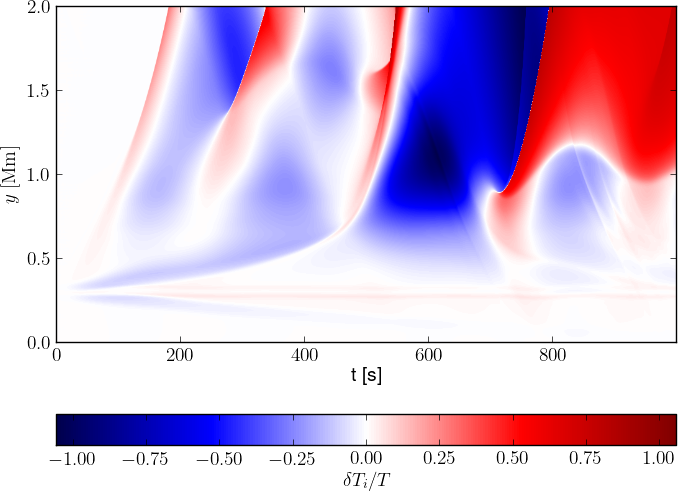}
        \hspace{0cm}
        \includegraphics[width=0.4\textwidth]{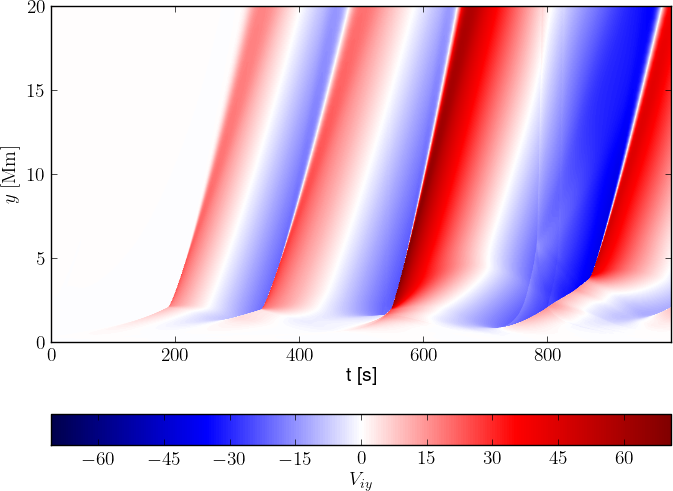}
        \hspace{-0cm}
        \includegraphics[width=0.4\textwidth]{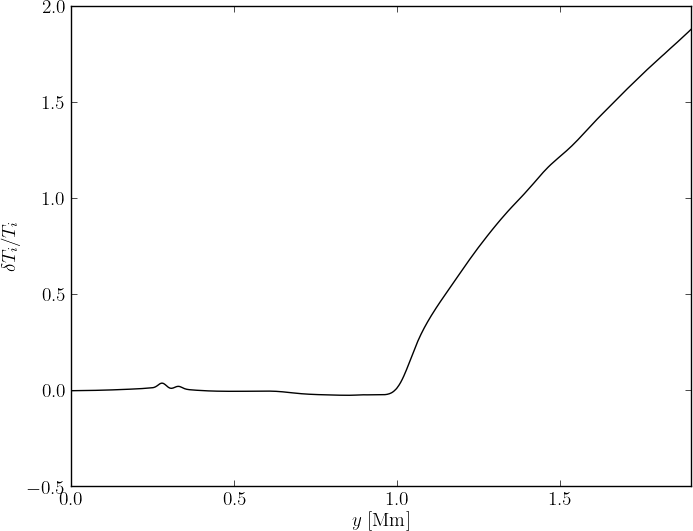}
        \hspace{0cm}
        \includegraphics[width=0.4\textwidth]{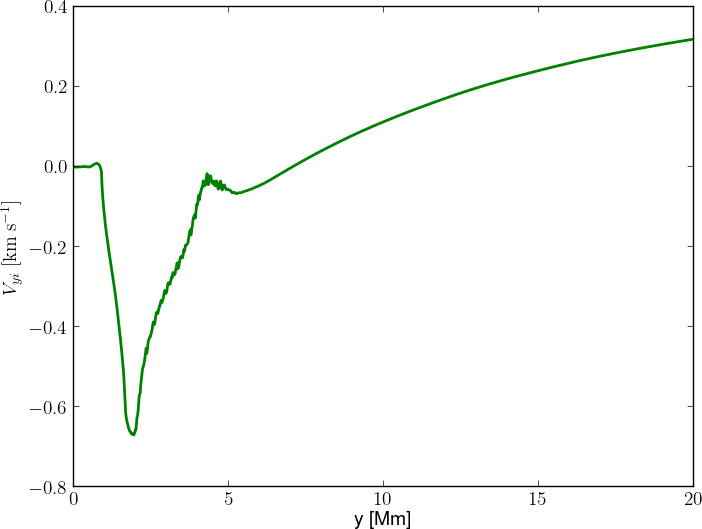}
    \end{center}
    \caption{Time-distance plots for $\delta T_{\rm i}/T$ (top-left) and $V_{\rm iy}$ (top-right), perturbed temperature of ions $\delta T_{\rm i}/T$ averaged over time (bottom-left), and the vertical component of ion velocity $V_{\rm iy}$ averaged over time (bottom-right) for $y_{\rm 0} = 0.3\;$Mm and $A=10\;\text{km}\cdot\text{s}^{-1}$.}
    \label{fig:maxViy_pulse_1_and_10}
\end{figure*}

Figure~5 (top panels) presents time--distance plots for the perturbed ion temperature $\delta T_{\rm i}/T$ (left panel) and for the vertical component of the ion velocity $V_{\rm iy}$ (right panel) in the case of $A=10\;\text{km}\cdot\text{s}^{-1}$ and $y_{\rm 0}=0.3\;$Mm (i.e.\ more or less in the middle of the photosphere). Some correlation between the velocity and temperature signals is discernible from this analysis. Mainly, in the range of $0\;$Mm$\;\le y \le 2\;$Mm the signal in $\delta T_{\rm i}/T$ experiences a similar trend to the signal in $V{\rm iy}$ with $\max(V_{\rm iy}) \approx 65\;\text{km}\cdot\text{s}^{-1}$ and $\max(\delta T_{\rm i}/T) \approx 1$ taking place for the launching level $y_{\rm 0} = 0.3\;$Mm. In the case of the same launching point $y_0$, but for a much smaller pulse amplitude, mainly $A = 1\;\text{km}\cdot\text{s}^{-1}$, the maximum value of the vertical component of ion velocity is almost 20 times lower than for this amplitude $A = 10\;\text{km}\cdot\text{s}^{-1}$ pulse. Similarly, the maximum value of the perturbed relative ion temperature is about 25 times lower for the simulation with amplitude $A = 1\;\text{km}\cdot\text{s}^{-1}$. As a matter of fact, for $A = 1\;\text{km}\cdot\text{s}^{-1}$ we obtained $\max(V_{\rm iy}) \approx 3.6\;\text{km}\cdot\text{s}^{-1}$ and $\max(\delta T_{\rm i}/T) \approx 0.05$ (not shown).
The bottom panels of Figure~5 illustrate the perturbed relative ion temperature $\delta T_{\rm i}/T$ averaged over time, and the vertical component of the ion velocity $V_{\rm iy}$ averaged over time. These two quantities are defined as
\begin{equation}
   H(y)=\frac{1}{t_{\rm 2}-t_{\rm 1}}\int_{t_{\rm 1}}^{t_{\rm 2}}\frac{\delta T_{\rm i}}{T}\, dt\, ,
\end{equation}
and
\begin{equation}
   V(y)=\frac{1}{t_{\rm 2}-t_{\rm 1}}\int_{t_{\rm 1}}^{t_{\rm 2}}V_{\rm iy}\, dt\, ,
\end{equation}
where $t_{\rm 1}=0\;$s and $t_{\rm 2}=5000\;$s. 
We note that the max($H(y)$) takes place for $y=2$ Mm (bottom-left). However, a small increase in temperature is discernible at $y \approx 0.3\;$Mm, the height at which the initial pulse is launched in this case. From the bottom-right panel of Fig.~5, it is inferred that a slow downflow takes place in the lower atmospheric layers up to $y \approx 5\;$Mm. Higher up, on the other hand, there is an upflow or outflow, with a magnitude that is growing with height $y$; it reaches about $0.35\;\text{km}\cdot\text{s}^{-1}$ at $y=20\;$Mm. Below $y \approx 5\;$Mm, the downflow reaches a minimum value of $V_{\rm iy} \approx -0.7\;\text{km}\cdot\text{s}^{-1}$. This means that when going from the lower atmosphere to the corona, the downflow turns into an upflow. A similar scenario was reported by \cite{2015SoPh..290.2889K} who found even faster downflows of $0 - 13\;\text{km}\cdot\text{s}^{-1}$ and also faster upflows of $10 - 12\;\text{km}\cdot\text{s}^{-1}$. Hence, our simple 2.5D numerical findings reveal only the general trend of the vertical flows that is also present in the observational data. This vertical flow can be generated by ponderomotive force originating from Alfv\'en waves  \citep[e.g.,][]{1992SoPh..139..279M}.

Figure~6 displays the relative perturbed ion temperature averaged over height and time. This quantity is defined as (with $H(y)$ given by Eq.~(23)):
\begin{equation}
   H=\frac{1}{y_{\rm 1}-y_{\rm 0}}\int_{y_{\rm 0}}^{y_{\rm 1}}H(y)\, dy\, ,
\end{equation}
where $y_{\rm 1}=20\;$Mm. The top panel illustrates $H$ versus the launching height $y_{\rm 0}$. We note that the minimal heating occurs for $y_{\rm 0} = 0\;$Mm and the maximal heating $H$ is obtained launching the pulse at $y_{\rm 0} = 0.35\;$Mm, which is close to the middle of the photosphere. We also note that $H$ strongly falls off with $y_{\rm 0}>0.35\;$Mm. Such behavior of $H$ versus $y_{\rm 0}$ results from the magnetoacoustic waves that release thermal energy as a result of the ion--neutral collisions and which are driven by Alfv\'en waves. The ion--neutral collisions are less frequent in the upper layers as the ionization degree is close to one (100\%) there.

Figure~6 (bottom panel) illustrates the dependence of the average heating $H$ on the pulse amplitude $A$ for a pulse located at $y_{\rm 0} = 0.25\;$Mm. Here, according to our expectations, it is clearly seen that the higher the amplitude of the pulse, the more heat is deposited in the atmosphere.
\begin{figure}
    \begin{center}
        \hspace{-0.2cm}
        \includegraphics[width=0.5\textwidth]{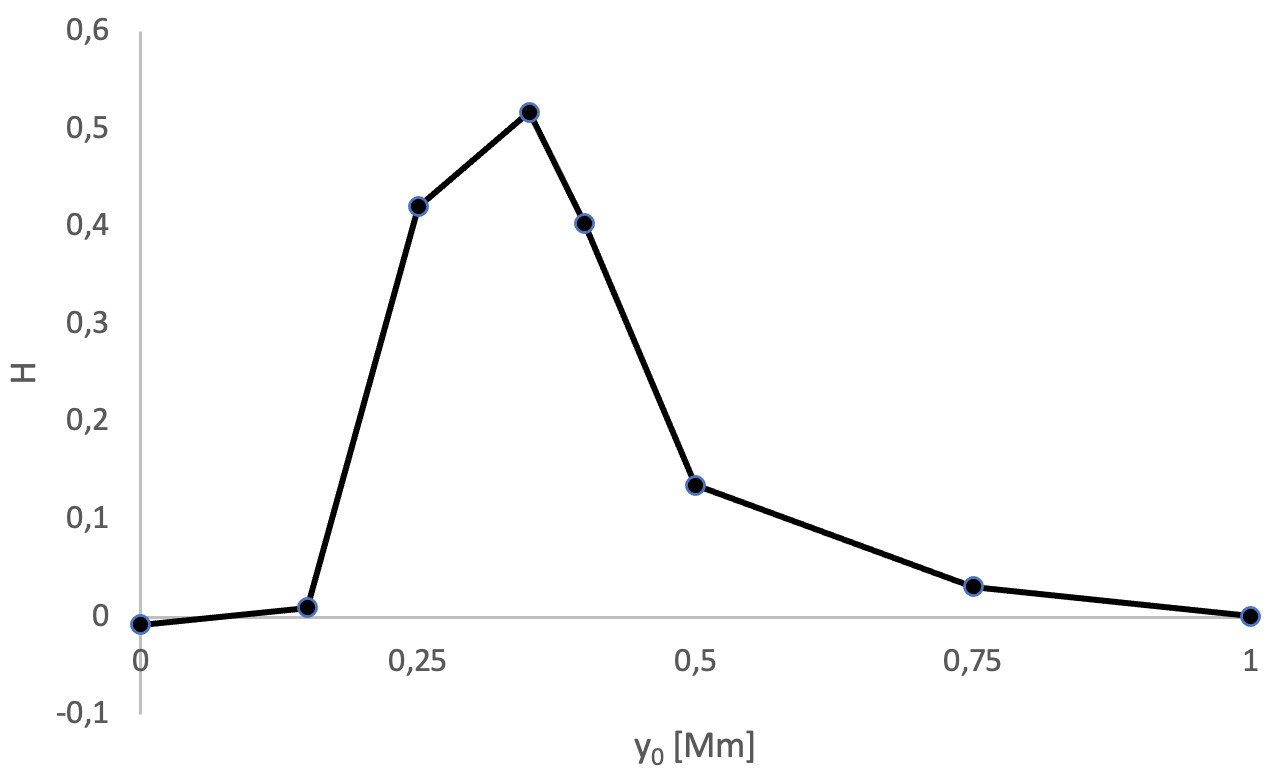}\\
        \vspace{0.5cm}
        \includegraphics[width=0.5\textwidth]{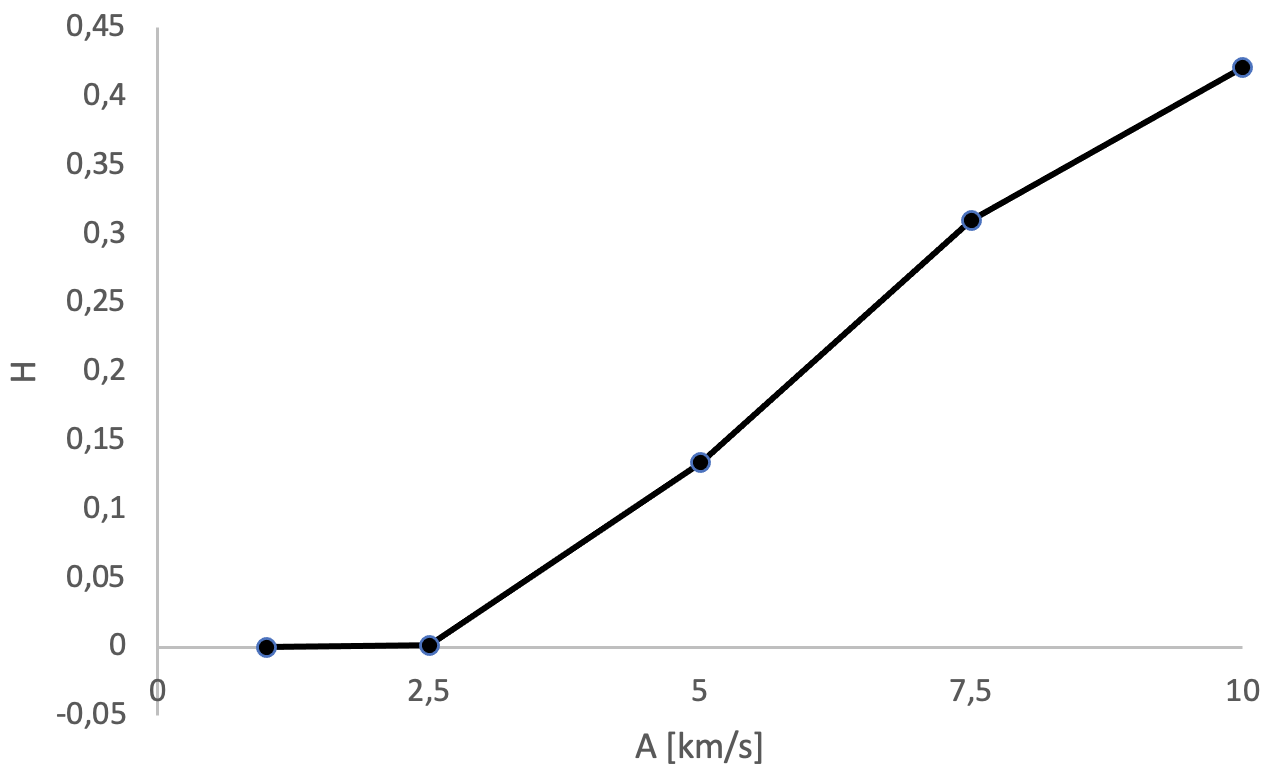}
    \end{center}
    \vspace{-0.25cm}
    \caption{Relative perturbed ion temperature averaged over time and height  $H$ vs.\ pulse launching $y_{\rm 0}$ for $A = 10\;\text{km}\cdot\text{s}^{-1}$ (top) and vs.\ pulse amplitude $A$ for $y_{\rm 0} = 0.25\;$Mm (bottom).}
    \label{fig:H}
\end{figure}
\section{Summary and Conclusion}
In this paper, we present the results of 2.5D simulations of impulsively generated two-fluid Alfv\'en waves. We show that as a consequence of ion--neutral collisions \citep[e.g.,][]{2018SSRv..214...58B}, large amplitude Alfv\'en waves generated in the photosphere and chromosphere can contribute to the heating of the chromosphere \citep[e.g.,][]{2020ApJ...896L...1M}. These waves can also drive plasma outflows, which higher up can become more substantial and constitute the origin of the solar wind. However, the magnitude of the flows is found to be rather moderate and much lower than the observed in- and outflows by \citet{2009ApJ...704..883T} and \citet{2015SoPh..290.2889K}. Hence, we conclude that impulsively generated two-fluid Alfv\'en waves with an initial amplitude $A = 10\;\text{km}\cdot\text{s}^{-1}$ are not able to explain the observational data, even though the obtained flow amplitudes are in the observed ranges. Nevertheless, the magnitude of the obtained plasma heating is proportional to the pulse amplitude. We find that the maximum heating occurs for the pulse launched from the middle of the photosphere. The maximum of the averaged relative temperature $H$ in the chromosphere increases by about 50\%. The pulses launched from the low photosphere and in the chromosphere do not substantially affect the chromosphere temperature (Fig.~6, top). Also, the magnitude of the obtained outflows grows with the amplitude of the pulse. In particular, in the case of the pulse launched from the middle of the photosphere and with an amplitude $A \le 2.5\;\text{km}\cdot\text{s}^{-1}$, the averaged heating $H$ is negligibly small. For larger values of the amplitude $A$, the trend in $H(A)$ is essentially linear with $H \approx 0.42$ for $A = 10\;\text{km}\cdot\text{s}^{-1}$ (Fig.~6, bottom).

Our model is based on the two-fluid equations in which the only source of plasma heating are ion--neutral collisions. Hence, more realistic models that include other nonideal and nonadiabatic effects are required to more accurately describe the solar atmosphere and make the simulations more realistic. Such models are left for future studies and the obtained results will be reported elsewhere. 

\section*{Acknowledgments}
The JOANNA code has been developed by Darek Wójcik. This work was done within the framework of the projects from the Polish National Foundation (NCN) grants Nos. 2017/25/B/ST9/00506 and 2020/37/B/ST9/00184. Numerical simulations were performed on the LUNAR cluster at Institute of Mathematics of University of M.\ Curie-Skłodowska, Lublin, Poland. SP acknowledges support from the project (EUHFORIA 2.0) taht has received funding from the European Union’s Horizon 2020 research and innovation programme under grant agreement No 870405. SP also received support from the projects
C14/19/089  (C1 project Internal Funds KU Leuven), G.0D07.19N  (FWO-Vlaanderen), SIDC Data Exploitation (ESA Prodex-12), and the Belspo projects BR/165/A2/CCSOM and B2/191/P1/SWiM. 

\bibliographystyle{aa}
\bibliography{main.bbl}
\end{document}